\documentclass[iop,superscriptaddress]{emulateapj}
\usepackage{amsmath,epsfig,url,natbib}
\usepackage{graphicx,epstopdf,float,color,array}
\usepackage{rotating,multirow}
\submitted{}
\shorttitle{The Quenching Slope}
\shortauthors{}

\definecolor{orange}{cmyk}{0,0.5,1,0}

\begin{document}
\title{Bridging between the integrated and resolved main sequence of star formation}

\author{Shoubaneh Hemmati \altaffilmark{1}, Bahram Mobasher\altaffilmark{2}, Hooshang Nayyeri\altaffilmark{3}, Abtin Shahidi\altaffilmark{2}, Peter Capak\altaffilmark{4}, Behnam Darvish\altaffilmark{4}, Nima Chartab\altaffilmark{2}, Marzyeh Jafariyazani\altaffilmark{2}, Zahra Sattari\altaffilmark{2}}
\email{shemmati@caltech.edu}

\altaffiltext{1}{Jet Propulsion Laboratory, California Institute of Technology, Pasadena, CA 91109, USA}
\altaffiltext{2}{University of California Riverside, 900 University Avenue, Riverside, CA 92521, USA}
\altaffiltext{3}{University of California Irvine, Irvine, CA 92697, USA}
\altaffiltext{4}{California Institute of Technology,1200 E California Blvd, Pasadena, CA 91125, USA}

\journalinfo{\textsuperscript{\textcopyright} 2019. All rights reserved. Accepted by Astrophysical Journal Letters.}
\begin{abstract}

The position of galaxies on the stellar mass, star formation rate plane with respect to the star-forming main sequence at each redshift is a convenient way to infer where the galaxy is in its evolution compared to the rest of the population. We use Hubble Space Telescope high resolution images in the GOODS-S field from the CANDELS survey and fit multi wavelength lights in resolution elements of galaxies with stellar population synthesis models. We then construct resolved kpc-scale stellar mass, star formation rate surface density curves for galaxies at $z\sim 1$. Fitting these resolved main sequence curves with Schechter functions we parameterize and explain the multi-wavelength structure of galaxies with three variables: $\phi^{*}$, $ \alpha$, and $M^{*}$. For quenched galaxies below the main sequence, we find an average high mass slope ($\alpha$) of the resolved main sequence curves to be $\sim -0.4$. The scatter of this slope is higher among the lower mass star forming galaxies and those above the main sequence compared to quenched galaxies, due to lack of an evolved bulge. Our findings agree well with an inside-out quenching of star-formation. We find that the knee of the Schechter fits ($M^{*}$) for galaxies below the main sequence occurs at lower stellar mass surface densities compared to star forming galaxies, which hints at how far quenching has proceeded outwards.
\end{abstract}


\section{Introduction}

The past thirteen billion years witnessed a vast evolution from tiny quantum fluctuations to the wide variety of galaxy structures and shapes we observe today. Studies of scaling relations among the physical properties of galaxies and their evolution with cosmic time have not only been critical in shaping our current understanding of galaxy formation and evolution (e.g., \citealt{Faber1976}, \citealt{Tully1977}), but are as well a magnificent aid in reducing systematics in cosmology (e.g., \citealt{Huff2014})

One such scaling relation, is the tight correlation between the star formation rate (SFR) and stellar mass in star-forming galaxies (i.e. star-forming main sequence) out to high redshifts (e.g., \citealt{Lilly2013}, \citealt{Speagle2014}, \citealt{Salmon2015}), with the bulk of star formation occurring primarily in massive star-forming galaxies as compared to less massive systems (e.g., \citealt{Noeske2007}, \citealt{Bell2005}, \citealt{Reddy2006}).  A population of passive galaxies also exists and lie below the main sequence, whereas starbursts lie above. The slope of the main sequence, is almost constant at each epoch but the zero point evolves significantly with time. The cause of the evolution in the main sequence is not very well understood. There is also discrepancy among the slope and scatter around the relation reported in different studies using different SFR diagnostics and various sample selections.

With the ever increasing wealth of high resolution observations,
we have, however, entered a new era that moves beyond studying galaxies as single data points and into one that systematically probes resolved properties of large samples of galaxies (e.g., \citealt{Wuyts2012},\citealt{Guo2012}, \citealt{Hemmati2014}, \citealt{Hemmati2015}, \citealt{Ellison2019}). In this paper, we explore the resolved star formation surface density vs. stellar mass surface density plane of $z\sim 1$ galaxies and its relation to the integrated star-forming main sequence. \S \ref{sec:data}, introduces the sample and data used in this work. In \S \ref{sec:resolvedms} we explain the method used to measure the resolved main sequence for individual galaxies (i.e., star formation surface density vs. stellar mass surface density). The bridge between the resolved main sequence and the integrated SFR-stellar mass plane is presented in \S \ref{sec:result}. Finally in \S \ref{sec:disc}, we address and discuss sources of uncertainty. Throughout this paper all magnitudes are expressed in AB system (\citealt{Oke1983}) and we use standard cosmology with $H_{0}=70 \ \rm km \ s^{-1}\  Mpc^{-1}$, $\Omega _{M}= 0.3$, and $\Omega _{\Lambda} = 0.7$.

\section{sample}\label{sec:data}

\begin{figure*} 
\centering
  \includegraphics[trim=0cm 0cm 0cm 0cm, clip,width=0.99 \textwidth]{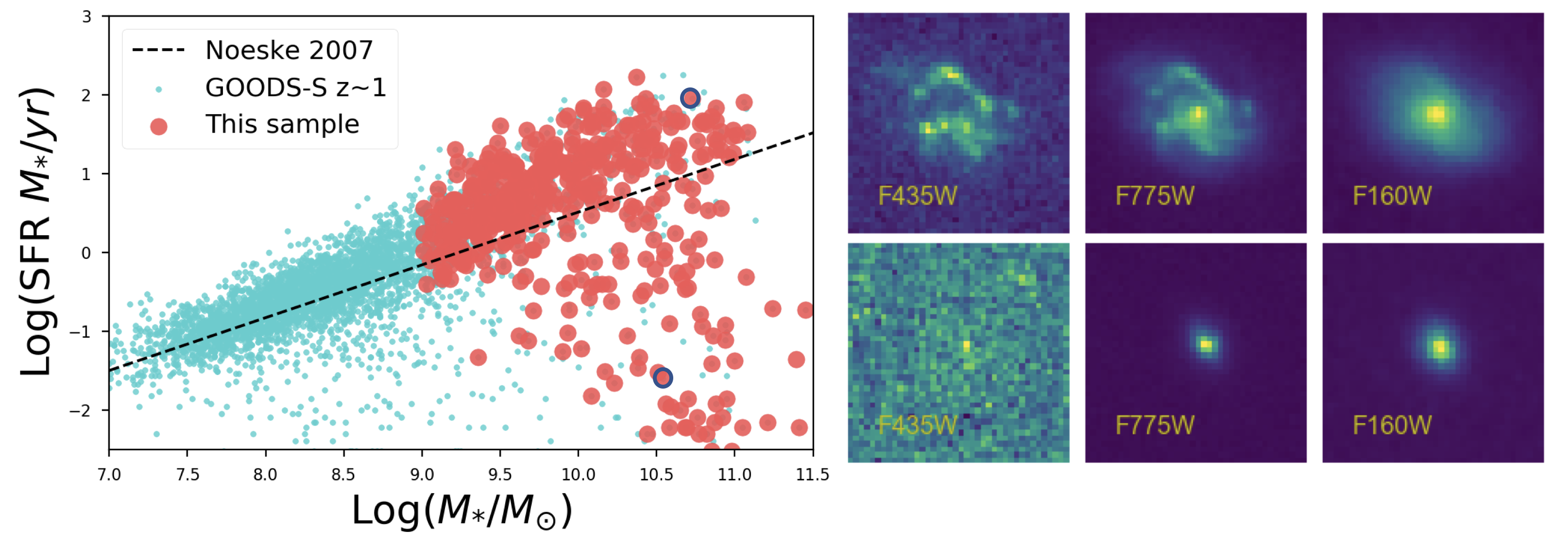}
\caption{Left:Galaxies selected for this study are shown as red circles on top of all GOODS-S galaxies at $z \sim 1$ shown with cyan dots. Black dashed line shows the main sequence of star forming galaxies at this redshift from \cite{Noeske2007}. Right: HST cutouts in F435W, F775W, and F160W of a star forming and quenched galaxy on top and bottom panels, respectively.}
\label{fig:sample}
\end{figure*}

Galaxies for this study are selected from the GOODS-S field where science-ready deep multi-band high resolution optical and near-IR HST images from the Cosmic Assembly Near-IR Deep Extragalactic Legacy Survey, CANDELS (\citealt{Grogin2011},\citealt{Koekemoer2011}) exist. We use images in four optical pass bands from the Advanced Camera for Survey: F435W, F606W, F775W, and F850lp as well as three from the Wide Field Camera 3 (WFC3): F105W, F125W, and F160W. 

CANDELS also publicly provides catalogs (\citealt{Guo2013} in the GOODS-S ) of multi-waveband photometries along with estimates of integrated physical properties of galaxies (e.g., redshifts, sizes, stellar masses, SFRs). We focus our studies here on $z \sim 1$ galaxies, where the HST observations cover rest-frame UV to near-IR wavelengths and hence, can provide reliable estimates of the resolved physical properties. This redshift regime is also optimal for our aim because while the star-forming main sequence of galaxies at $z \sim 1$ is well in place and well studied there are also plenty of galaxies below the main sequence to study statistically.

In the left panel of Figure \ref{fig:sample}, we present the sample selected for this work on top of all $z\sim 1$ galaxies in the GOODS-S field on the star formation rate-stellar mass plane. Both stellar mass and star formation rate here are taken from the public CANDELS catalogs. Beside our selection cut of stellar mass ($>10.0^{9}$ $M_{*}/M_{\odot}$) and redshift ($0.8<z<1$), we also remove elongated galaxies ($b/a>0.5$) and those fainter than $26$ AB mag in the F160W and we only include galaxies with spectroscopic redshifts. This leaves us with 545 galaxies. We show HST cutouts in F435W, F775W and F160W filters for two galaxies (a star forming galaxy on top and a passive galaxy on the bottom) from the sample in the right panels of Figure \ref{fig:sample}.

\section{resolved main sequence}\label{sec:resolvedms}

To measure resolved properties for the galaxies in our sample from the HST images, we follow the same steps outlined in \citealt{Hemmati2014}, using an improved and now fully automated python pipeline (see e.g., \citealt{Jafariyazani2019}). The pipeline makes large $100 \times 100$ pixel cutouts from the seven HST bands centered on the RA and DEC of each sample galaxy, where each pixel in the image corresponds to $0.06"$. The resolution of the shorter wavelength cutouts are then matched to the resolution of F160W, using the convolution method in \textit{astropy} and a matching kernel created by the \textit{photutils} package using the PSFs of HST images. The initial large size of the cutouts is selected to remove issues due to PSF matching near edges of the cutouts. After PSF matching smaller cutouts of $40\times 40$ pixels are regenerated, which correspond to a physical scale of $\sim 20$ kpc at $z\sim 1$. To make the pixel photometric catalog per galaxy we mask background pixels surrounding the galaxy using a segmentation routine which we wrote and fine tune to find edges of galaxies. We used image processing modules from the \textit{SciPy.ndimage} in the segmentation maker routine. 

\begin{figure*} 
\centering
  \includegraphics[trim=0cm 0cm 0cm 0cm, clip,width=0.99 \textwidth]{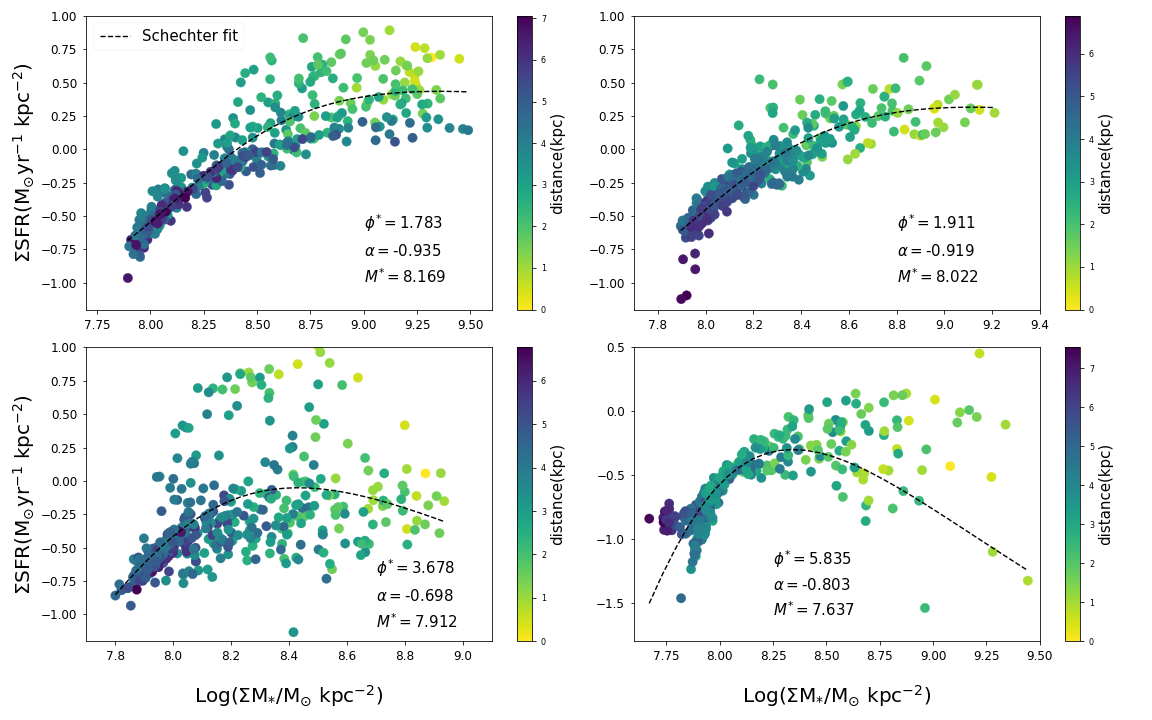}
\caption{Resolved star formation surface density vs. stellar mass surface density plots of four sample galaxies color coded by distance from the center. Dashed black lines show the best Schechter function fit to data.}
\label{fig:resolved_ms}
\end{figure*}

Given the pixel based photometric catalogs, we fit for the resolved kpc-scale physical properties at each resolution element with the LePhare (\citealt{Arnouts1999}, \citealt{Ilbert2006}) SED fitting code. We use the Bruzual and Charlot (\citealt{Bruzual2003}) stellar population synthesis models, a Chabrier initial mass function (\citealt{Chabrier2003}), exponentially declining or $\tau$ model star formation histories, sub-solar stellar metallicity, and the Calzetti dust attenuation curve (\citealt{Calzetti2000}) to build the library of model SEDs. The justification for these choices in building the model library and the effect of modifications to these assumptions is fully discussed in \citealt{Hemmati2014}. There, we also performed tests of accuracy of the resolved SED fitting measurements by comparing integrated and sum of resolved measured values. 

The best fitting model representing resolved galaxy photometries is found and used by LePhare to assign stellar mass surface densities and SFR surface densities to individual resolution elements in all galaxies in our sample. Figure \ref{fig:resolved_ms} shows the SFR surface density versus the stellar mass surface density or the so-called resolved main sequence of star formation for four sample galaxies color-coded with distance of the resolution element from the center of each galaxy. 

The shapes of the resolved main sequence curves are very informative about the physical properties of galaxies. Stellar mass surface density increases towards smaller distances from the center of the galaxy. SFR surface density also mostly increases from the outer disks towards the center but with decreasing slopes. It is important to keep in mind that both stellar mass and SFR surface densities are fit with the same SED fitting code and models and hence the tightness of these curves can be partly due to the dependence of both on the library of model SEDs used to fit for these physical properties. However, each resolution element is fit independently of the others and the tight distributions observed are mostly due to gradual change of photometry from one resolution element to the next. 

The general shape of the resolved main sequence is rising in low stellar mass densities, reaches a peak, and falls towards massive pixels located in the center of galaxies. This shape can be parameterized with a Schechter function \eqref{eq:1} which is typically used to fit the luminosity or mass function of galaxies. 

\begin{equation}
SF(m_{*}) = \phi^{*} \times 10^{(M^{*} - m_{*})(\alpha + 1)} \times e^{-10^{(M^{*} - m_{*})}}  
\label{eq:1}
\end{equation}

In equation \eqref{eq:1}, $SF$ and $m_{*}$ are the star formation rate and stellar mass surface densities respectively. We used \textit{scipy.curvefit} with bounds on possible values to fit for the three free parameters: $\phi^{*}$ the normalization factor, $\alpha$ the high mass slope, and $M^{*}$ the characteristic mass or the knee of the resolved main sequence. The best fit Schechter functions and their parameters are also shown in Figure \ref{fig:resolved_ms} for the four sample galaxies.

\section{The bridge}\label{sec:result}

Location on the integrated SFR-$M_{*}$ plane with respect to the main sequence of star forming galaxies at each redshift is a simple way to infer how far each individual galaxy is in its evolution. For instance, \citealt{Scoville2017}, based on multi-variable model fitting bins galaxies out to high redshift on integrated SFR-$M_{*}$ planes and studies star formation efficiencies and interstellar medium mass estimates from ALMA sub-mm observations as a function of distance from the star-forming main sequence. Here, we adopt a similar strategy and bin our sample on the integrated SFR-$M_{*}$ plane to study how the shape of the resolved main sequence curves vary across this plane. 

\begin{figure*} 
\centering
  \includegraphics[trim=5cm 2cm 5cm 1cm, clip,width=0.99\textwidth]{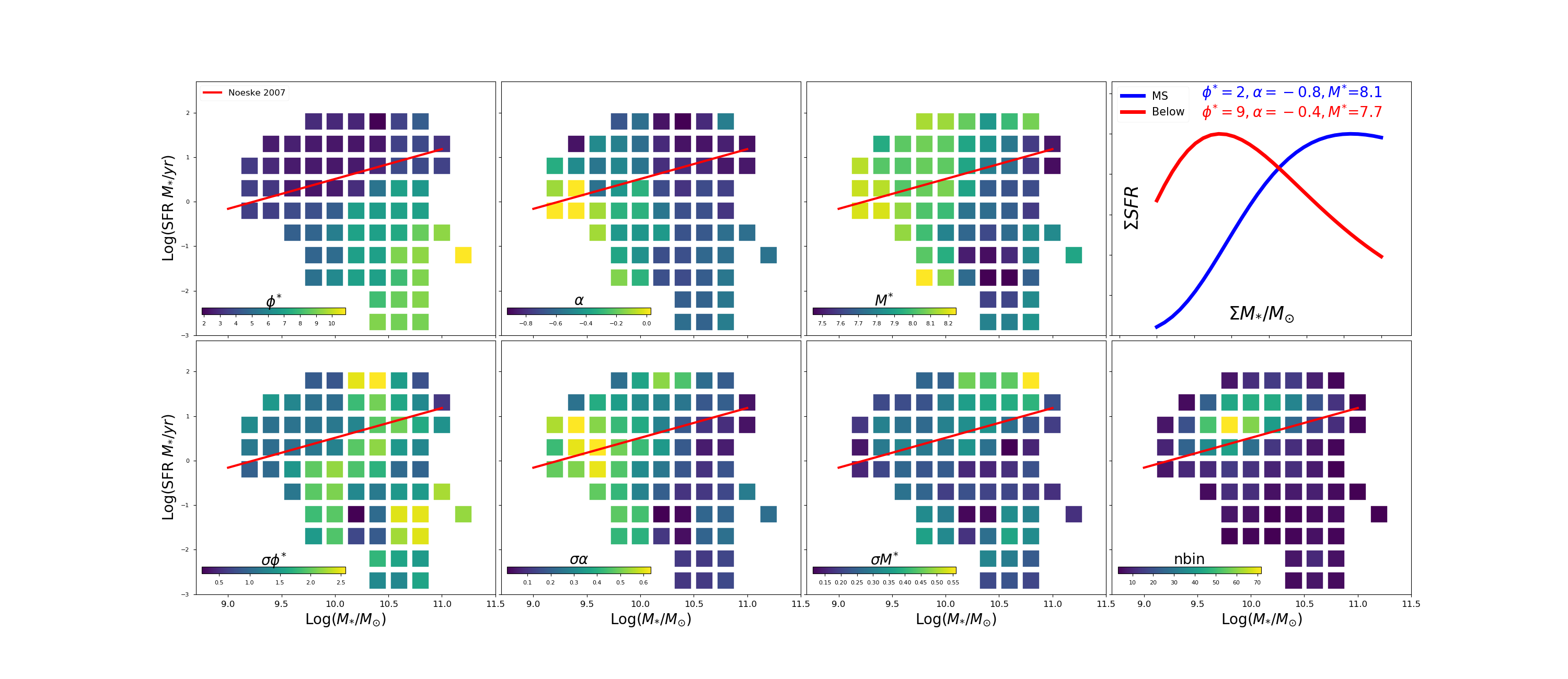}
\caption{Schechter parameters best fitted to the resolved main sequence of star formation of individual galaxies are shown on the binned integrated stellar mass-SFR plane. Left-three columns show the median (top) and standard deviation (bottom) of $\phi^{*}$, $\alpha$, and $M^{*}$. Red solid line shows the main sequence of star-forming galaxies at $z\sim 1$ to guide the eye. Top-right panel shows two Schechter functions with average value of galaxies on and two dex below the main sequence at $Log(M_{*}/M_{\odot})\sim 10.5$. Bottom-right panel shows number of galaxies in each bin of SFR and stellar mass.}
\label{fig:params}
\end{figure*}

The left three columns of Figure \ref{fig:params} show the median (top) and standard deviation (bottom) of best fit Schechter parameters in bins of stellar mass and SFR. We have excluded galaxies with poor Schechter fits. We calculated the goodness of fit using the diagonal elements of the covariance matrix of parameters measured by $\it curvefit$. We also exclude bins with only one galaxy. Right-bottom panel of Figure \ref{fig:params} shows the number of galaxies in each bin. The integrated star-forming main sequence at $z \sim 1$ (\citealt{Noeske2007}) is over plotted on all panels (red solid line).  

The median of $\phi^{*}$ is lower for galaxies on and above the star-forming main sequence compared to the galaxies below the star-forming main sequence.  $\phi^{*}$ increases smoothly at each fixed stellar mass bin with decreasing SFR and at fixed SFR bins, with increasing stellar mass. The standard deviation of $\phi^{*}$ however does not show dependence on location with respect to the star-forming main sequence. 

 The variation in $\alpha$ values is not as clear as $\phi^{*}$. As can be seen from the $\sigma \alpha$ plot the variation is lower in the higher mass end of the main sequence compared to lower mass bins. This can be explained by the evolution of the central bulge of galaxies towards higher masses. More massive galaxies have more evolved bulges, with highest stellar mass surface densities. By definition evolved bulges have lower specific star formation rates compared to a less evolved bulge. $\alpha$ or the high stellar mass surface density slope of the Schechter function corresponds to the spread of the pixels in the central parts of the galaxy where the bulges form. If most resolution elements have high stellar mass and low SFR surface densities, scatter of alpha is low and if some pixels are still forming stars at higher rates then the bulge is not fully evolved and scatter is large. Our results show that the scatter of $\alpha$ is low at higher mass galaxies, meaning the central parts of these galaxies are more settled compared to lower mass galaxies, where either the bulge has not fully formed yet or the star formation is still on going in some of the resolution elements. The presence of an evolved bulge imposes strong constraint on the high mass end of the Schechter functions. The high $\sigma \alpha$ values above the main sequence, also point to the increased scatter in their resolved main sequence curves at higher stellar mass surface densities, which in addition to lack of an evolved bulge can be possibly due to presence of more star forming clumps in these systems. At fixed stellar mass bins however, an increase in $\alpha$ is seen as SFR decreases. 

Third column of Figure \ref{fig:params} shows variation of $M^{*}$ across the bins of stellar mass and SFR. Clear decrease of $M^{*}$ diagonally from low mass and high SFR to high mass and low SFR distance indicates that the Schechter function breaks in lower characteristic masses in more massive and evolved systems. The average scatter of $M^{*}$ in bins is $\sim 0.3$ dex and the maximum reaches 0.5 dex in the bin with the highest integrated stellar mass and SFR.

The top-right panel of Figure \ref{fig:params}, summarizes the plot by showing an average shape (with normalized y axis) of the resolved main sequence curve for galaxies (at $Log(M_{*}/M_{\odot})\sim 10.5$) on and below the main-sequence.

\section{Discussion}\label{sec:disc}

Comparison of the Schechter fits to resolved main sequence curves on and below the integrated star-forming main sequence points towards an inside-out quenching of star formation. The more quiescent galaxies below the main sequence have their knee of the Schechter fit at lower masses and quenching is pushed towards outer radii. Our findings in this paper conforms well with previous literature linking the inside-out quenching to the growth of the central bulge (e.g., \citealt{Fang2013}, \citealt{Hemmati2015P}, \citealt{Tacchella2015}, \citealt{Morselli2019}, \citealt{Jafariyazani2019}).

Here, we used the Schechter function to fit the resolved main sequence curves for the first time. Similar to the literature on luminosity functions one concern can be why Schechter functions are chosen and not double power laws or other functional forms. Here, the aim was to investigate whether and how the shape of the resolved main sequence informs us about its integrated properties, specifically its position with regards to the main sequence of star formation and whether it is quenched. Schechter function parameters, i.e., the faint end slope (here the high mass slope), the knee or the characteristic luminosity/mass/magnitude, and the normalization are well known by the astronomy community and very easy to visually interpret. 

We have also excluded sources with poor Schechter fits from the final analysis. Apart from possibility of intrinsically distinct structural profiles in a small subset of targets (e.g., AGNs, mergers), large errors in some of the Schechter fits can be due to imperfections/failures in various steps of the automated measurement (e.g., segmentation mask, resolved SED-fitting on low surface brightness resolution elements). Here, the sample is being trimmed based on curve fitting error rather than in the beginning based on stellar mass surface density and S/N of pixels. Choosing the cut value is not straight forward. Having a low fitting error threshold ($ E < 0.25$) results in removing $80\%$ of objects, with the majority of the remaining galaxies being massive and star forming. Changing this threshold to a very large value ($E<20$) results in removing only $10\%$ of objects from the sample $98\%$ of which are less massive than $log(M_{*}/M{\odot}) =9.75$. $E<5$ removes $18\%$ of galaxies as poor fits with $97\%$ of them being less massive than 9.75. We chose galaxies with ($E<1.5$) removing $30\%$ of the sample, $92\%$ of which being galaxies with $log(M_{*}/M_{\odot}) <9.75$. The exact value of the cut is arbitrary, however with the chosen value the fits visually represent the overall spread of data on the $M_{*}-SFR$ plane well. Moreover, the trends in Figure \ref{fig:params} are there regardless of this choice. The trends only get clearer where better fits are kept and having a moderate value of $1.5$ ensures enough statistics.

The resolved main sequence curves also clarify to some extent, the scatter around the main sequence of star formation. Given which galaxies end up in a given sample, the more bulge dominated ones will bring the integrated measure down while the star forming clumps within a disk will push the whole thing up. We note here again the dependence of the stellar mass and SFR surface density on the model library used for fitting the SED of each resolution element. The continuity of the shapes seen in the resolved curves, depend on parameter steps used in defining the model library. This is because the photometry is changing rather smoothly in kpc-scale at $z\sim1$ galaxies with changing radius. One may use an independent measurement of the resolved SFR to improve upon this analysis. This can be done using Hydrogen Balmer lines observed with integral field spectrographs to measure dust corrected SFR from H$\alpha$ and H$\beta$. Future work is also needed to extend this analysis to a broader redshift range and study the evolution of star formation in galaxies. The high spatial resolution and longer wavelength coverage of the James Webb Space telescope makes it an optimal choice for this purpose.

We wish to thank the referee for providing very useful suggestions and recommendations. S.H. is thankful to A. Pagul for careful reading of this manuscript and for constructive comments. B.D. acknowledges financial support from the National Science Foundation, grant number $1716907$. Parts of this research were carried out at the Jet Propulsion Laboratory, California Institute of Technology, under a contract with the National Aeronautics and Space Administration.

\bibliography{quench.bib}
\end{document}